# Micropolarity Ramification of Asymmetric Merging Flow

**Abuzar Abid Siddiqui**
Department of Basic Sciences, College of Engineering, B. Z. University, Multan-60800, Pakistan.

**ABSTRACT**

*The steady, asymmetric, and two-dimensional flow of viscous, an incompressible micropolar fluid through a rectangular channel with a splitter (parallel to walls) was formulated and simulated numerically. The geometric parameter that controls the position of splitter was defined as splitter position parameter. The plane Poiseuille flow was considered far from upstream and downstream of the splitter. A numerical scheme that comprises a fourth order method followed by special finite-difference method was used to solve the boundary value problem. This numerical scheme transforms the governing equations to system of finite–difference equations which we have solved by S.O.R. iterative method. Moreover, the results obtained were further refined and upgraded by the Richardson extrapolation method. The results were compared on different grid sizes as well as with the existing results for symmetric flow of Newtonian fluids. The comparisons were satisfactory. The microrotation effects on the splitter plate were significantly high as compared to other locations of the channel. This highest value for the asymmetric channel was higher in magnitude as compared to for the symmetric one. Furthermore, the proclivity of rotation of aciculate elements of the fluid about their centroids rises if either the Reynolds' number increases or/and the distance between the upper and splitter plates decreases.*

## 1. INTRODUCTION

If two oncoming channels flows are combined to constitute a single channel flow then it is named as merging flow [1-2]. It has a lot of applications not only in the external flows like lakes, rivers and estuaries [3-4], but also in the internal flows like internal machinery dynamics, respiratory flow [5-6] and quadrupole magnetic cell sorter [7]. Earlier, this type of flow was investigated symmetrically [8-12] as well as for asymmetrically for the Newtonian fluids. In this study, we examined, numerically, a steady, two-dimensional asymmetric merging flow of the *micropolar (Cosserat) fluids*. The micropolar fluids are the generalized Newtonian fluids having aciculate rod-like or dumbbell shaped molecules which can spin about their centroids. These fluids are exemplified by blood [13-14], polymers and polymeric suspension [15], rigid-rod epoxies [16], colloidal suspension and liquid crystals [17]. These fluids are being used on lab-on-a-chip [18-19] nowadays. These fluids differ from Newtonian fluids in two ways. These fluids can sustain couple stress and body couples. Secondly, the shear stress tensor is non–symmetric. On the other hand, micropolar fluids are also not non–Newtonian fluids because they exhibit the micro–inertia due to the spin of the colloids or molecules of the fluids.

Motivated by the micropolar fluids which have techno-scientific importance, we decided to study the micropolarity effect on the fluid flowing in asymmetric rectangular channel. Our interest lies in the cross-sectional distribution of flow lines, longitudinal fluid speed, microrotation of the colloidal particles of the fluids, and the skin friction. In addition, this analysis comprises the effect of the movement of the splitter plate within the channel as well as that of Reynolds number on the bulk flow. The graphical forms of the results obtained are presented in section 5. The formulation and basic flow analysis are presented in section 2 while sections 3 and 4 contain the detail of numerical scheme so adopted and computational procedure used respectively.

## 2. BASIC ANALYSIS

Let us write the governing equations, neglecting thermal effects, of an incompressible viscous micropolar fluid [17], in dimensional form, as:

$$\mathbf{V}'_{n,n} = 0, \tag{1}$$

$$\rho \dot{\mathbf{V}}'_k = \rho \mathbf{b}'_k + \sigma'_{jk,j}, \tag{2}$$

and
$$\rho j_o \dot{\mathbf{N}}'_k = \rho \mathbf{c}'_k + \mathbf{m}'_{jk,j} + \varepsilon_{kjn}\sigma'_{jn}. \tag{3}$$

In addition, its constitutive equations are [20-21]

$$\sigma'_{kj} = -p\delta_{kj} + (\mu+\chi)(\mathbf{V}'_{j,k} + \mathbf{V}'_{k,j}) + \chi(\mathbf{V}'_{k,j} - \varepsilon_{kjn}\mathbf{N}'_n), \tag{4}$$

and
$$\mathbf{m}'_{kj} = a_1 \mathbf{N}'_{n,n}\delta_{kj} + a_2 \mathbf{N}'_{k,j} + a_3 \mathbf{N}'_{j,k}. \tag{5}$$

Here the superscript dot signifies for the material derivative such that $\mathbf{V}'$ and $\mathbf{N}'$ are, respectively, are the (dimensional) fluid-velocity and the microspin of the colloids in the fluid; $\mathbf{b}'$ and $\mathbf{c}'$ are, respectively, are the body-force and the body-couple; $p$, $\rho$ and $j_o$ are, respectively, the fluid-pressure, the fluid-density and the local micro (colloidal)-inertia [13]. Apart three spin-gradient- viscosity-coefficients $\{a_1, a_2, a_3\}$ while $\{\lambda, \mu, \chi\}$ are three spin-viscosity-coefficients. These are related as the following traditional form of the Clausius-Duhem inequality, viz.: [21]

$$\left.\begin{array}{l} 3a_1 + a_2 + a_3 \geq 0, \ a_3 \geq a_2, \ a_3 \geq 0 \\ 3\lambda + 2\mu + \chi \geq 0, \ 2\mu \geq -\chi, \ \chi \geq 0 \end{array}\right\} \tag{6}$$

The mechanics of the problem under consideration can briefly be stated as: The flow is through an infinite length non-uniform channel $|y'| \leq h, x' \in (-\infty, \infty)$ with finite width $2h$ parallel walls. The upstream region is divided into two channels with the help of semi-infinite splitter plate while downstream is flow between two plates without any plate. The walls and splitter are parallel to each other and x-axis. The position of splitter is controlled by a splitter position parameter α. If α=0 then the splitter plate will coincide with the x–axis while the plate will lie above and below the x-axis if α>0 and α<0 respectively, as shown in figure 1. Moreover, the flow is considered as steady, two-dimensional, laminar, and asymmetric about x-axis in general. The flow is generated by the uniform pressure gradient analogous to that of plane Poiseuille flow on inlet and outlet flow to preserve the continuity.

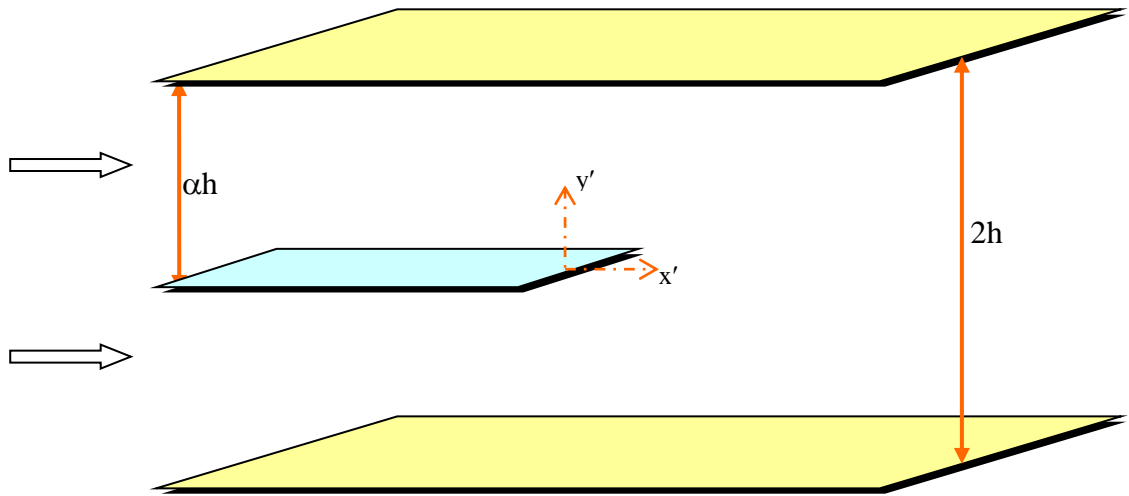

Figure 1. The flow configuration.

On introducing the dimension analysis, viz.:

$$\mathbf{V}'_i = \mathbf{V}_i U_\infty, \quad x'_i = h x_i, \quad \omega' = \omega \frac{U_\infty}{h} \text{ and } \mathbf{N_i} = \frac{U_\infty}{h} \mathbf{N}_i,$$

such that $\omega' = \nabla' \times \mathbf{V}$, in the light of the above assumptions equations (1)-(3) in components form will deform as

$$\frac{\partial u}{\partial x} + \frac{\partial v}{\partial y} = 0, \tag{7}$$

$$\nabla^2 E - C_1 \nabla^2 \xi = R\left[u\frac{\partial E}{\partial x} + v\frac{\partial E}{\partial y}\right], \tag{8}$$

and

$$\nabla^2 \xi - C_2[E\text{-}2\xi] = C_3\left[u\frac{\partial \xi}{\partial x} + v\frac{\partial \xi}{\partial y}\right], \tag{9}$$

where $u$, $v$ are the $x$ and $y$ components of velocity vector respectively while $E$ and $\xi$ are components of vorticity and spin respectively, in addition to three dimensionless constants $\{C_1, C_2, C_3\} = \left\{\frac{\chi}{\mu+\chi}, \frac{\chi h^2}{\gamma}, \frac{\rho U_\infty j h}{\gamma}\right\}$ and the Reynolds number $R = \frac{\rho U_\infty h}{\mu+\chi}$. Next, equations (7)–(9), in stream function $\psi$ form, will yield

$$E = -\nabla^2 \psi, \tag{10}$$

$$\nabla^2 E - \frac{C_1}{2}\nabla^2 \eta = R\left[\frac{\partial \psi}{\partial y}\frac{\partial E}{\partial x} - \frac{\partial \psi}{\partial x}\frac{\partial E}{\partial y}\right], \tag{11}$$

and

$$\nabla^2 \eta - C_2[E\text{-}\eta] = \frac{C_3}{2}\left[\frac{\partial \psi}{\partial y}\frac{\partial \eta}{\partial x} - \frac{\partial \psi}{\partial x}\frac{\partial \eta}{\partial y}\right]. \tag{12}$$

Here $\eta = 2\xi$.

For the geometry of figure 1 the boundary conditions for these equations are as follows:

(a) No slip at the walls, i.e. u=v=0, on all the solid walls and the splitter plate. (13a)

(b)
$$\left.\begin{array}{l} \psi \to \dfrac{1}{(1-\alpha)^3}\left[-2y^3 + 3y^2 + 3\alpha(y^2 - 2y + \alpha) - \alpha^3\right] \\[2mm] E \to -\dfrac{1}{(1-\alpha)^3}[12y - 6 - 6\alpha] \end{array}\right\} \text{ as } x \to -\infty \tag{13b}$$

(c) $\psi \to 1.5y - 0.5y^3$ and $E \to 3y$ as $x \to \infty$ (13c)

Here equations (13b) and (13c) are due to imposed Poiseuille flow far upstream and far downstream of the splitter. Note that if we put α=0 in the boundary conditions given in (2.6) then it will become boundary condition of symmetric flow. To solve the above boundary value problem, we use the numerical scheme whose description is given in section 3, and which reduces the highly non-linear system of partial differential equations to system of difference equations. This system is further solved by the S. O. R. - iterative method, which has capability to accelerate the convergence of the iterative scheme. Henceforth this numerical procedure is efficient, for studies of such type of complex flow problems and is straightforward, economical in core storage requirements of a computer, and easy to programme.

## 3. THE NUMERICAL SCHEME

This numerical scheme consists of two methods. The grid size along $x$ and $y$ directions, will be taken by $H$, $K_1$ respectively, and around a typical internal grid point $(x_0, y_0)$ we adopt the convention that quantities at $(x_0, y_0)$, $(x_0+H, y_0)$, $(x_0, y_0+K_1)$, $(x_0-H, y_0)$, $(x_0, y_0-K_1)$, $(x_0+2H, y_0)$, $(x_0, y_0+2K_1)$, $(x_0-2H, y_0)$, and $(x_0, y_0-2K_1)$ will be represented by the subscripts 0, 1, 2, 3, 4, 5, 6, 7, and 8 respectively.

### Fourth order Finite-Difference Method

If we employ the method similar to as that of [22] then governing equations (10)-(12) yield:

$$\frac{1}{12H^2}\{16\psi_1 + 16\psi_3 - \psi_5 - \psi_7\} + \frac{1}{12K_1^2}\{16\psi_2 + 16\psi_4 - \psi_6 - \psi_8\} - \frac{5}{2}\left\{\frac{1}{H^2} + \frac{1}{K_1^2}\right\}\psi_0 = -E_0, \quad (14)$$

$$\left(\frac{4}{3H^2} + 8P_0\right)E_1 + \left(\frac{4}{3K_1^2} + 8Q_0\right)E_2 + \left(\frac{4}{3H^2} - 8P_0\right)E_3 + \left(\frac{4}{3K_1^2} - 8Q_0\right)E_4 - \left(\frac{1}{12H^2} + P_0\right)E_5$$

$$-\left(\frac{1}{12K_1^2} + Q_0\right)E_6 - \left(\frac{1}{12H^2} - P_0\right)E_7 - \left(\frac{1}{12K_1^2} - Q_0\right)E_8 - \frac{5}{4}\left(\frac{1}{H^2} + \frac{1}{K_1^2}\right)[2E_0 - \eta_0] \quad (15)$$

$$-\frac{2C_1}{3H^2}[\eta_1 + \eta_3] - \frac{2C_1}{3K_1^2}[\eta_2 + \eta_4] - \frac{C_1}{24H^2}[\eta_5 + \eta_7] - \frac{C_1}{24K_1^2}[\eta_6 + \eta_8] = 0,$$

and

$$\left(\frac{4}{3H^2} + S_0\right)\eta_1 + \left(\frac{4}{3K_1^2} + 8T_0\right)\eta_2 + \left(\frac{4}{3H^2} - 8S_0\right)\eta_3 + \left(\frac{4}{3K_1^2} - 8T_0\right)\eta_4 - \left(\frac{1}{12H^2} + S_0\right)\eta_5$$

$$-\left(\frac{1}{12K_1^2} + T_0\right)\eta_6 - \left(\frac{1}{12H^2} - S_0\right)\eta_7 - \left(\frac{1}{12K_1^2} - T_0\right)\eta_8 - C_2[E_0 - \eta_0] = 0,$$

(16)

where $P_0 = \dfrac{-R}{144HK_1}(-\psi_6 + 8\psi_2 - 8\psi_4 + \psi_8)$,  $Q_0 = \dfrac{R}{144HK_1}(-\psi_5 + 8\psi_1 - 8\psi_3 + \psi_7)$

$S_0 = \dfrac{C_3 P_0}{2R}$, and $T_0 = \dfrac{C_3 P_0}{2R}$.

### Special Finite-Difference Method

In order to approximate equation (10) by second order standard central difference formulation, at point "0", we get

$$\frac{1}{H^2}\psi_1 + \frac{1}{K_1^2}\psi_2 + \frac{1}{H^2}\psi_3 + \frac{1}{K_1^2}\psi_4 - \left\{\frac{2}{H^2} + \frac{2}{K_1^2}\right\}\psi_0 = -E_0 \quad (17)$$

and if we employ 4$^{th}$ order central difference formulations, at the point "0", we get,

$$\frac{1}{12H^2}\{16\psi_1 + 16\psi_3 - \psi_5 - \psi_7\} + \frac{1}{12K_1^2}\{16\psi_2 + 16\psi_4 - \psi_6 - \psi_8\} - \frac{5}{2}\left\{\frac{1}{H^2} + \frac{1}{K_1^2}\right\}\psi_0 = -E_0 \quad (18)$$

Now equations (11)-(12) will be transformed into finite-difference equations by the special finite-difference method. Let us split equation (11) into following four equations:

$$\frac{\partial^2 E}{\partial x^2} + B\frac{\partial E}{\partial x} = A, \tag{19}$$

$$\frac{\partial^2 E}{\partial y^2} + C\frac{\partial E}{\partial y} = -\frac{1}{3}A, \tag{20}$$

$$D\frac{\partial^2 \eta}{\partial x^2} = -\frac{1}{3}A, \tag{21}$$

$$D\frac{\partial^2 \eta}{\partial y^2} = -\frac{1}{3}A, \tag{22}$$

where $A(x, y)$ is an unknown function; $B = -R\frac{\partial \psi}{\partial y}$, $C = R\frac{\partial \psi}{\partial x}$ and $D = -0.5C_1$.

Equation (19) is then approximated along the grid line $y=y_0$, over the range $x_0-H \leq x \leq x_0+H$, by applying a local transformation for E viz.

$$E = \lambda e^{-F}, \tag{23}$$

where
$$F = \frac{1}{2}\int_{x_0}^{x} B(t, y)dt \tag{24}$$

On inserting equation (23) in equation (19) and approximating the derivatives involved by second order central differences, we get

$$\lambda_1 + \lambda_3 - 2\lambda_0 - \left[\frac{1}{2}\left(\frac{\partial B}{\partial x}\right)_0 + \frac{1}{4}B_0^2\right]\lambda_0 H^2 = A_0 H^2 \tag{25}$$

Similarly, equation (20) is approximated along the grid line $x=x_0$, over the range $y_0-K_1 \leq y \leq y_0+K_1$, by applying a local transformation for $E$, given by

$$E = \mu e^{-G}, \tag{26}$$

where
$$G = \frac{1}{2}\int_{y_0}^{y} C(s, t)dt. \tag{27}$$

On using equation (26) in equation (20) after simplification, we have

$$\frac{H^2}{K_1^2}[\mu_2 + \mu_4 - 2\mu_0] - \left[\frac{1}{2}\left(\frac{\partial C}{\partial y}\right)_0 + \frac{1}{4}C_0^2\right]\mu_0 H^2 = -\frac{A_0}{3}H^2 \tag{28}$$

(3.14)

Now if we approximate equations (21) and (22) by second order central difference approximation we get,

$$D[\eta_1 + \eta_3 - \eta_0] = -\frac{A_0 H^2}{3}, \tag{29}$$

$$\frac{H^2 D}{K_1^2}[\eta_2 + \eta_4 - \eta_0] = -\frac{A_0 H^2}{3}. \tag{30}$$

On adding equations (25) and (28)–(30) and using the argument $\lambda_0 = \mu_0 = E_0$, we get,

$$\lambda_1 + \lambda_3 - 2E_0 - \frac{1}{4}E_0 B_0^2 H^2 + \left[\frac{H^2}{K_1^2}(\mu_2 + \mu_4 - 2E_0) - \frac{1}{4}E_0 C_0^2 H^2\right]$$
$$+ D[\eta_1 + \eta_3 - 2\eta_0] + \frac{DH^2}{K_1^2}[\eta_2 + \eta_4 - 2\eta_0] = 0 \tag{31}$$

In order to express $\lambda$ and $\mu$ back in terms of $E$, it can be done from the definitions. It is found that,

$$\lambda_i = E_i e^{F_i}, \qquad \mu_j = E_j e^{G_j}, \qquad (32)$$

where
$$F_i = \frac{1}{2}\int_{x_0}^{x_i} B(t, y_0)dt, \quad G_j = \frac{1}{2}\int_{y_0}^{y_j} C(x_0, t)dt, \qquad (33)$$

such that $i=1, 3$ and $j=2, 4$. Now we can replace $\lambda_1, \lambda_3, \mu_2, \mu_4$, by expressions involving $E_1, E_3, E_2, E_4$ respectively, but they will involve exponential coefficients. Next we expand the above exponentials in powers of their arguments keeping the truncation error of order $H^4$ and $H^2 K_1^2$. On expanding the exponents, we neglect the terms of order $H^4$, $H^2 K_1^2$ and their higher order. After some simplifications under above arguments and using the Taylor's theorem, we obtain:

$$\lambda_1 + \lambda_3 = [1 + \frac{H^2}{4}(\frac{\partial B}{\partial x})_0 + \frac{B_0^2 H^2}{8}][E_1 + E_3] + \frac{B_0 H}{2}[E_1 - E_3], \qquad (34)$$

$$\mu_2 + \mu_4 = [1 + \frac{K_1^2}{4}(\frac{\partial C}{\partial y})_0 + \frac{C_0^2 K_1^2}{8}][E_2 + E_4] + \frac{C_0 K_1}{2}[E_2 - E_4]. \qquad (35)$$

On Inserting equations (34) and (35) into the equation (31), we get,

$$\left[1 + \frac{B_0 H}{2} + \frac{B_0^2 H^2}{8}\right]E_1 + \left[\frac{H^2}{K_1^2}\{1 + \frac{K_1^2 C_0^2}{8}\} + \frac{C_0 H^2}{2K_1}\right]E_2 + \left[1 - \frac{B_0 H}{2} + \frac{B_0^2 H^2}{8}\right]E_3$$

$$+ \left[\frac{H^2}{K_1^2}\{1 + \frac{K_1^2 C_0^2}{8}\} - \frac{C_0 H^2}{2K_1}\right]E_4 + D\xi_1 + \left[\frac{DH^2}{K_1^2}\right]\eta_2 + D\eta_3 \qquad (36)$$

$$+ \left[\frac{DH^2}{K_1^2}\right]\eta_4 - \left[2D + \frac{2DH^2}{K_1^2}\right]\eta_0 - \left[2 + \frac{2H^2}{K_1^2} + \frac{B_0^2 H^2}{4} + \frac{C_0^2 H^2}{4}\right]E_0 = 0.$$

Next, by following a similar procedure for equation (3.7) as for equation (3.6), we get

$$\left[1 + \frac{U_0 H}{2} + \frac{U_0^2 H^2}{8}\right]\eta_1 + \left[\frac{H^2}{K_1^2}\{1 + \frac{K_1^2 V_0^2}{8}\} + \frac{V_0 H^2}{2K_1}\right]\eta_2 + \left[1 - \frac{U_0 H}{2} + \frac{U_0^2 H^2}{8}\right]\eta_3$$

$$+ \left[\frac{H^2}{K_1^2}\{1 + \frac{K_1^2 V_0^2}{8}\} - \frac{V_0 H^2}{2K_1}\right]\eta_4 + 2C_2 H^2 E_0 \qquad (37)$$

$$- \left[2 + \frac{2H^2}{K_1^2} + \frac{U_0^2 H^2}{4} + \frac{V_0^2 H^2}{4} + 2C_2 H^2 e^{2s_0}\right]\eta_0 = 0,$$

where $U = -c_3 \frac{\partial \psi}{\partial y}, \quad V = c_3 \frac{\partial \psi}{\partial x}$.

This approximation gives rise to an associated matrix that is always diagonally dominant and is of fourth order in spatial coordinates. Finally the condition for E is required at cylinder surface. Here we use the same approximation as that is done originally to Wood [23] and is given by

$$E_{w_1} = \frac{3}{K_1^2}[1 - \psi_a] - \frac{1}{2}E_a, \qquad \text{on the upper wall} \qquad (38a)$$

$$E_{w_2} = -\frac{3}{K_1^2}[1 + \psi_b] - \frac{1}{2}E_b, \qquad \text{on the lower wall} \qquad (38b)$$

$$E_{w_3} = -\frac{3}{K_1^2}\psi_c - \frac{1}{2}E_c, \qquad \text{on the upper wall} \qquad (38c)$$

where the subscripts $w_1$, $w_2$, $w_3$, represent for the value at the approximate boundary points on upper, lower, and splitter plate respectively while the subscripts $a$, $b$, and $c$ signify to the internal grid point most immediate to $w_1$, $w_2$, $w_3$ respectively.

## 4. **COMPUTATIONAL PROCEDURE**

This numerical experiment consists of two steps. In first step, Equations (17), (36) and (37) are solved iteratively with respect to the boundary conditions given in (13) and (38) by SOR-method [24] until the convergence is achieved according to the criterion:

$$\max\left|\eta_0^{(m+1)} - \eta_0^{(m)}\right| < 10^{-5}, \quad \max\left|E_0^{(m+1)} - E_0^{(m)}\right| < 10^{-5}, \quad \text{and} \quad \max\left|\psi_0^{(m+1)} - \psi_0^{(m)}\right| < 10^{-5} \quad (39)$$

In the second step, the equations (3.1)–(3.2) are solved numerically with respect to boundary conditions by SOR-method on using the data obtained from step 1 as an initial estimation. This procedure is repeated until the convergence is attained according the criterion given in (39). The solutions obtained are further extrapolated by Richardson's extrapolation technique [25].

## 5. **RESULTS, COMMENTS AND COMPARISONS**

The numerical calculations are carried out on grid sizes $H \in \{1/80, 1/160, 1/320\}$ and $K_1 \in \{1/120, 1/240, 1/480\}$ but the results presented are extrapolated. The range of the Reynolds number R and the splitter positions parameter $\alpha$ are taken as $R \in [0, 10^3]$ and $\alpha \in (-1, 1)$, respectively, but the representative values taken for display are $R \in \{1, 10^2, 10^3\}$ and $\alpha \in \{-0.5, 0, 0.2, 0.5\}$. Moreover the normalized coefficients $\{C_1, C_2, C_3\} = \{0.95, 0.0075, 1\}$, whereas the infinite length of the channel is numerically approximated and is taken as four throughout the study.

The streamlines of the flow in the channel for different values of R and $\alpha$ are presented in figures 2–4. These figures indicate that, for low Reynolds' number flow, the growth of the vortices below the splitter plate has been observed. These vortex effects increases as the splitter plate is moved from y=0 to y=1. Furthermore, these vortices diffuse as R increases as shown in figures 3 and 4.

Figures 5–7 are displayed for the equispinlines for different values of R and $\alpha$. These figures show that the maximum value of $|\eta|$ occurs on the splitter plate for all R. In addition, it increases as R increases for all $\alpha$. It also intensifies if the splitter plate moves towards upper plate. Moreover, this maximum value of $|\eta|$ for symmetric case ($\alpha=0$) is less than that for asymmetric flow case (when $\alpha \neq 0$). That explains that the proclivity of rotation of aciculate elements of the fluid about their centroids if either the Reynolds' number increases or/and the distance between the upper and splitter plates decreases.

In order to assess the micropolarity effects over Newtonian fluids, we examined the variation of $u_\alpha$ and $E_{up}$ along x–axis. Here $u_\alpha$ is the component of fluid–velocity along x–direction on the region between the trailing edge of the splitter and the right numerical approximated boundary of the channel, whereas $E_{up}$ is the vorticity on the upper plate. As for a representative set of values of R=1 and $\alpha \in \{-0.5, 0, 0.2, 0.5\}$ are presented in figures 8–9. These figures indicate that the highest value of $u_\alpha$ occurs in the vicinity of the trailing edge of the splitter for all $\alpha$ for both micropolar and Newtonian fluids. This highest value of $u_\alpha$ in micropolar fluids is greater than that in Newtonian ones. Analogous to this, the maximum value of $|E_{up}|$ in micropolar fluids is greater than that in Newtonian fluids. Moreover, when $\alpha=0$ it is observed that the velocity along splitter plate increases in downstream of the trailing edge of the splitter and then becomes uniform with x, but it is not so when α≠0. In addition, this trend is common in Newtonian as well as in micropolar fluids.

The vortex effects or in other words the variation of the skin friction ($C_f=-E$) on the upper, splitter, and lower plates of the channel are also examined for micropolar fluids. These are displayed in figures 10–12 for different values of R and α. The highest magnitude of the skin friction on either plate (upper, splitter, or lower) occurs in the upstream of the splitter plates. In addition it appears either flow is asymmetric or symmetric with respect to the splitter plate as well as for all values of Reynolds' number R. Furthermore, this highest magnitude of skin friction $C_f$ intensifies with increasing |α| for all R. It also increases as R increases for all α, regardless of the plates.

## 6. **CONCLUSION**

We formulated and examined numerically the steady asymmetric merging flow of the micropolar fluid in a rectangular channel having upstream splitter plate. The micropolar fluids comprise the colloids that can rotate about their centroids. This microspin affects the bulk flow. This effect was highlighted in this study. The microrotation effects on the splitter plate were significantly high as compared to other locations of the channel. This highest value for the asymmetric channel was higher in magnitude as compared to for the symmetric one. Moreover, the proclivity of rotation of aciculate elements of the fluid about their centroids rises if either the Reynolds' number increases or/and the distance between the upper and splitter plates decreases.

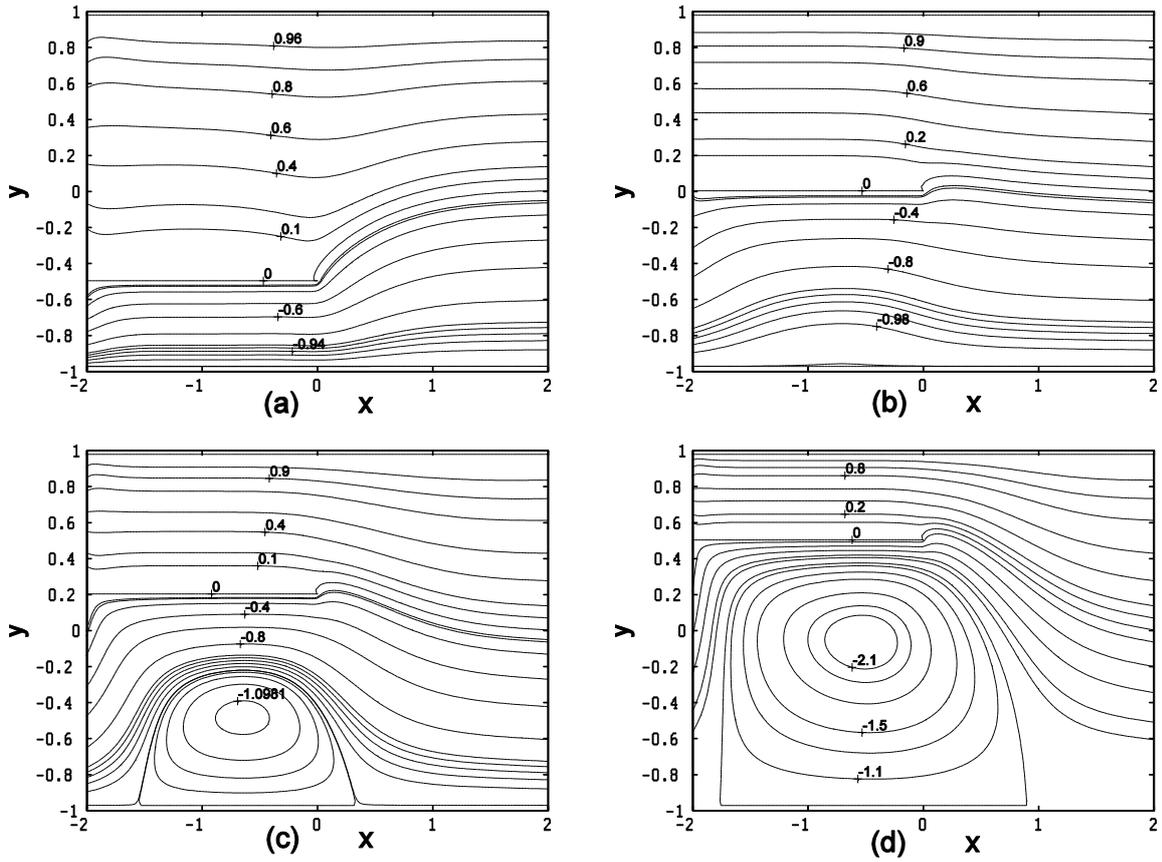

Figure 2. Streamlines for R=1, (a) α=-0.5, (b) α=0, (c) α=0.2, (d) α=0.5.

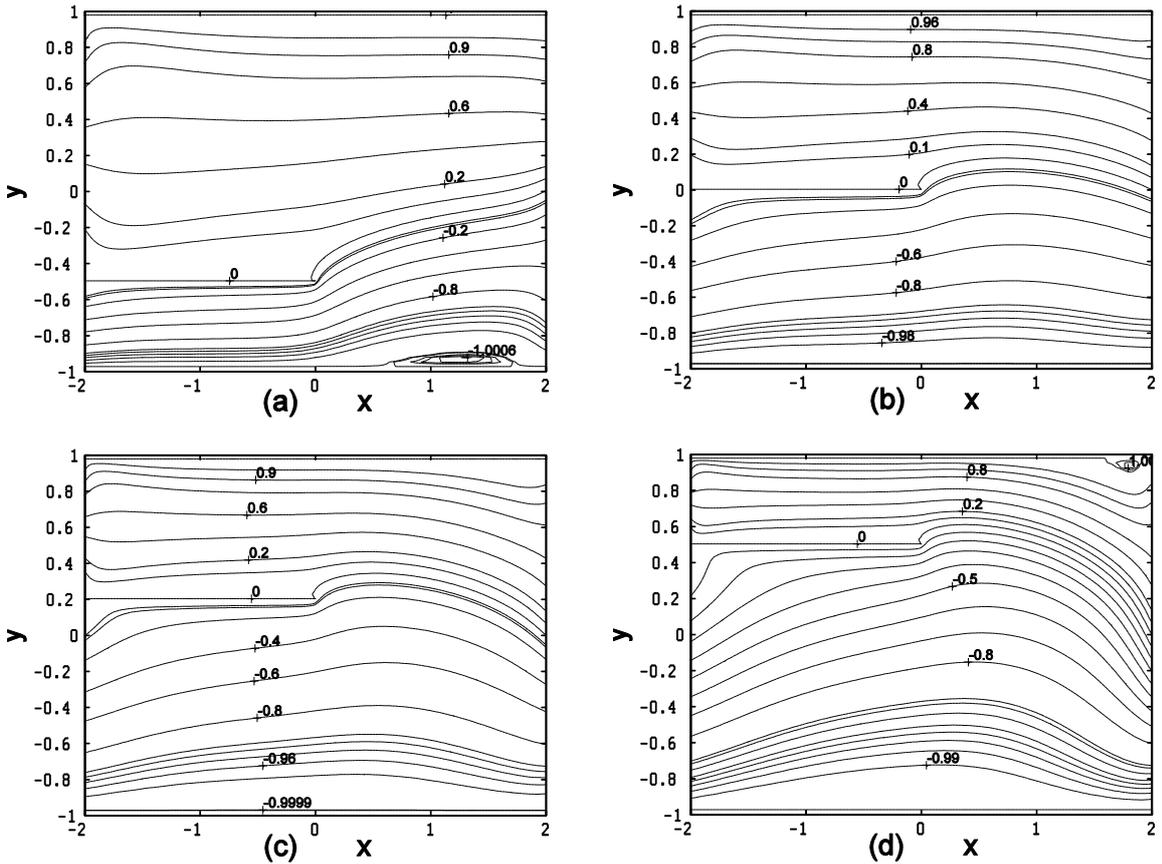

Figure 3. Streamlines for R=100, (a) α=-0.5, (b) α=0, (c) α=0.2, (d) α=0.5.

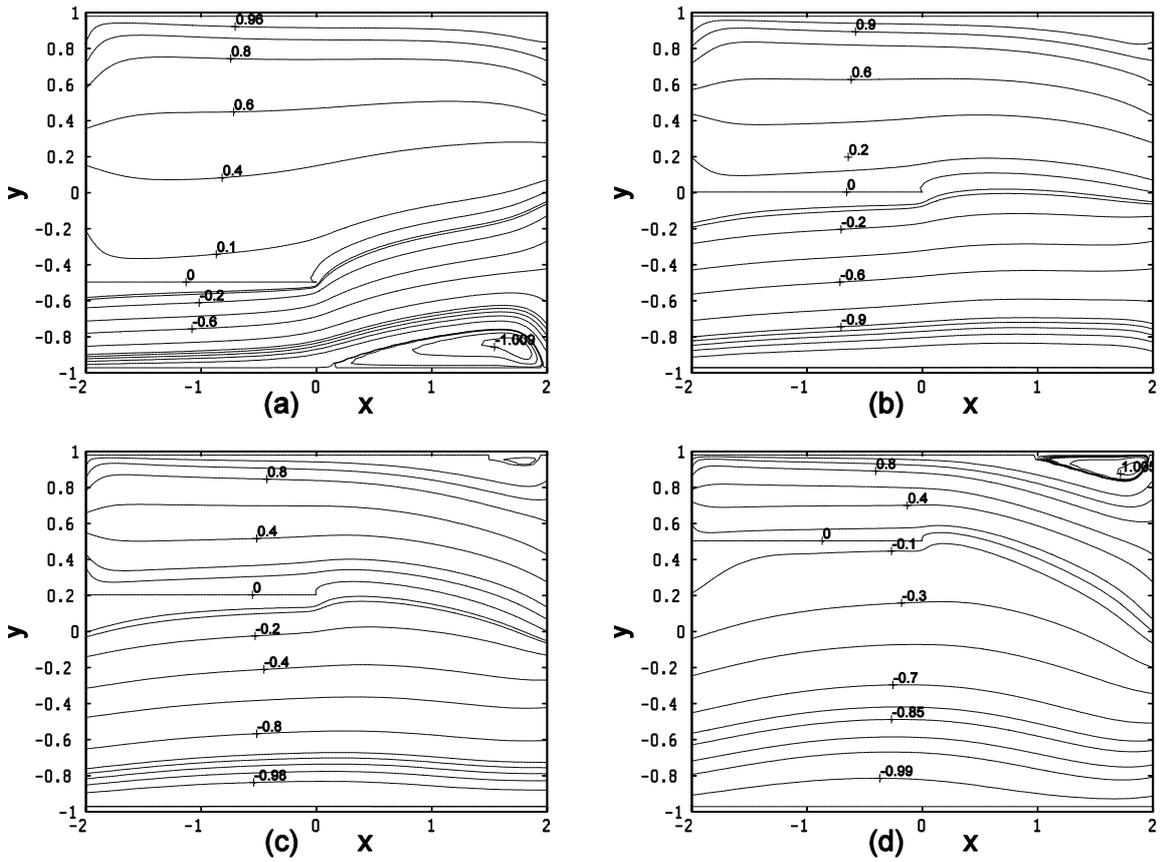

Figure 4. Streamlines for R=1000, (a) α=-0.5, (b) α=0, (c) α=0.2, (d) α=0.5.

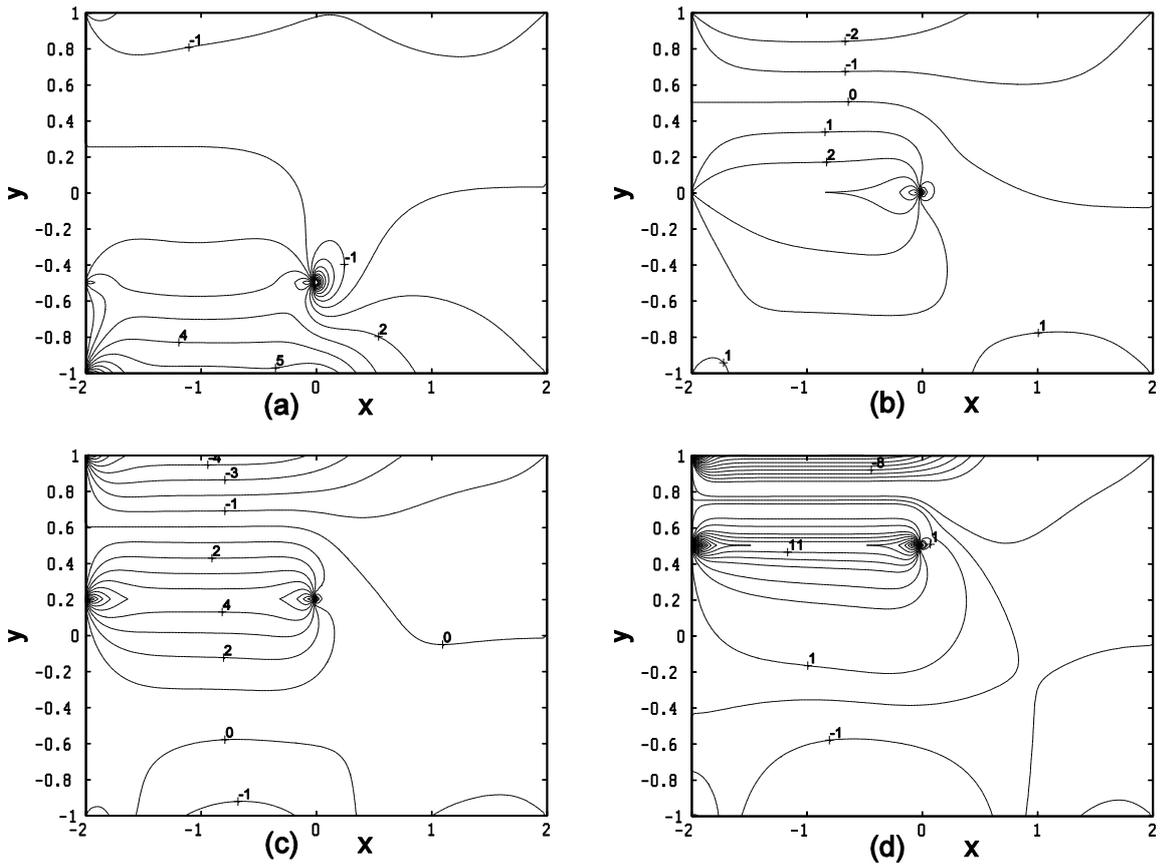

Figure 5. Iso-spinlines R=1, (a) α=-0.5, (b) α=0, (c) α=0.2, (d) α=0.5.

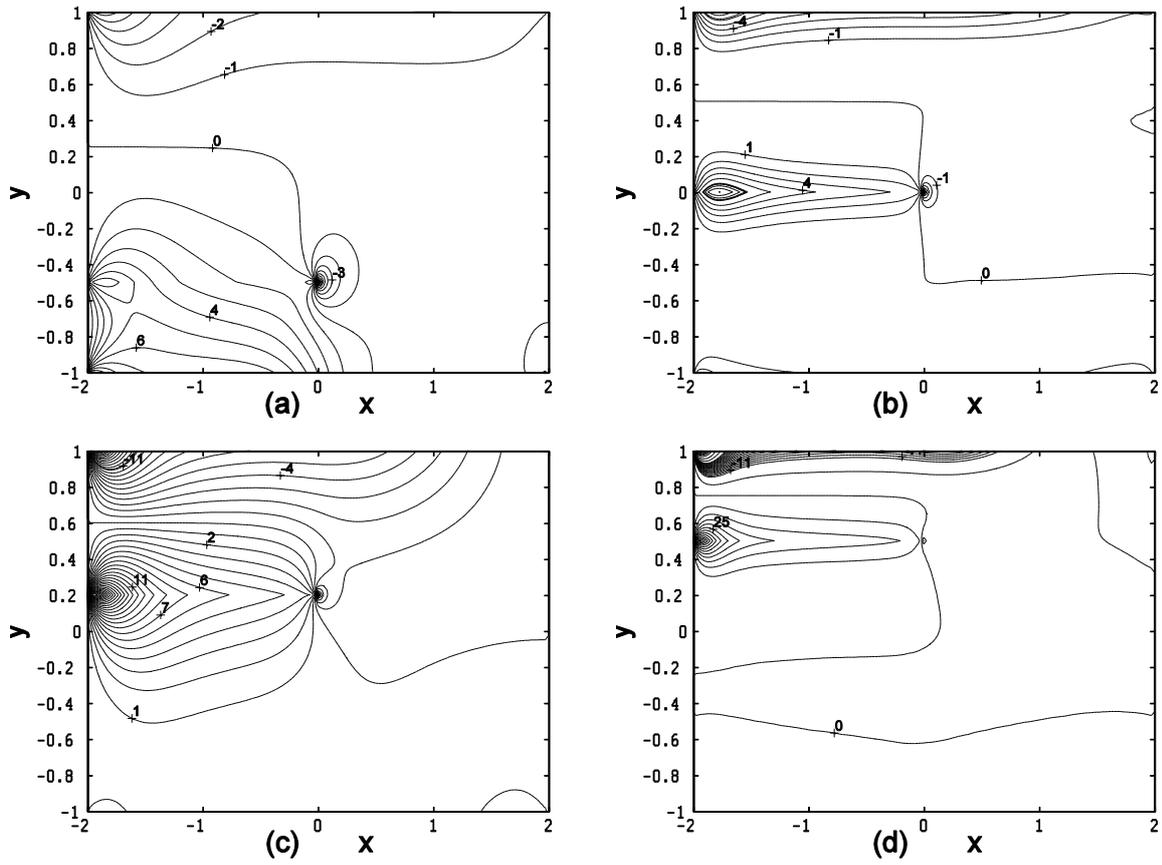

Figure 6. Iso-spinlines for R=100, (a) $\alpha=-0.5$, (b) $\alpha=0$, (c) $\alpha=0.2$, (d) $\alpha=0.5$.

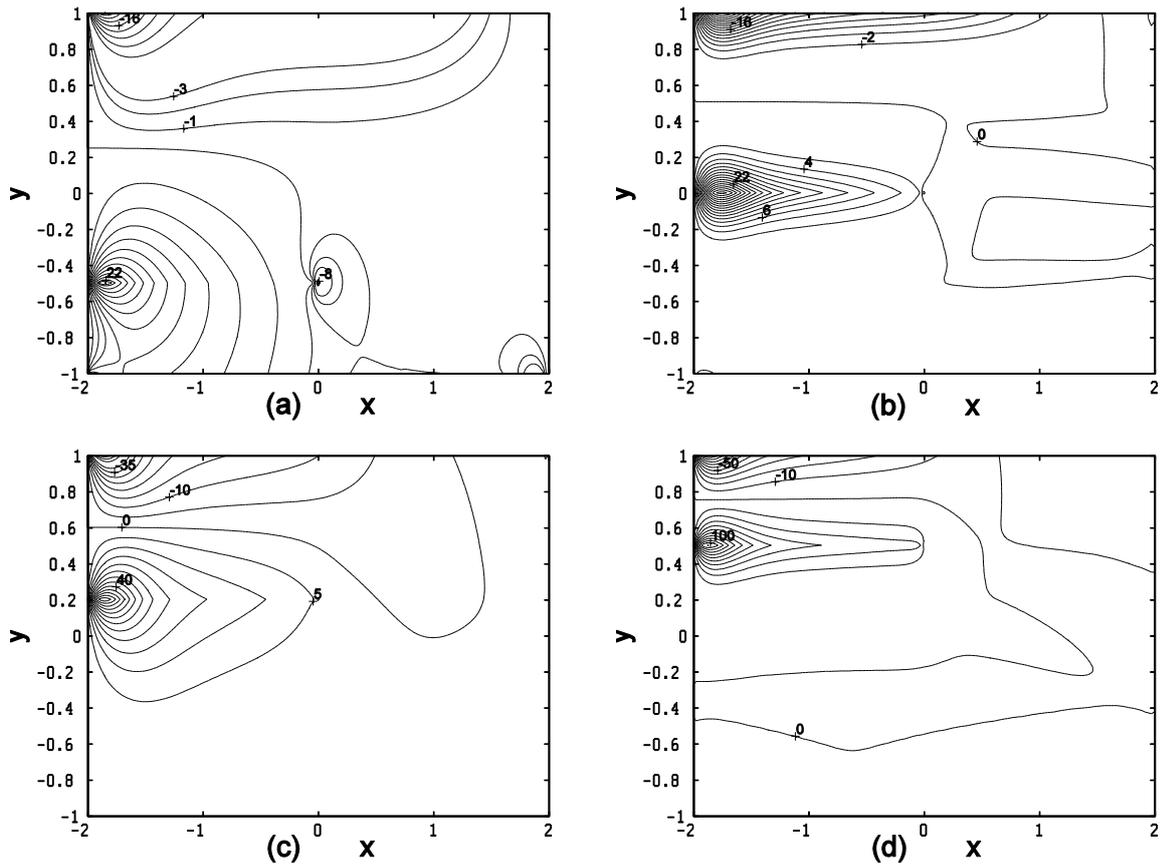

Figure 7. Iso-spinlines for R=1000, (a) $\alpha=-0.5$, (b) $\alpha=0$, (c) $\alpha=0.2$, (d) $\alpha=0.5$.

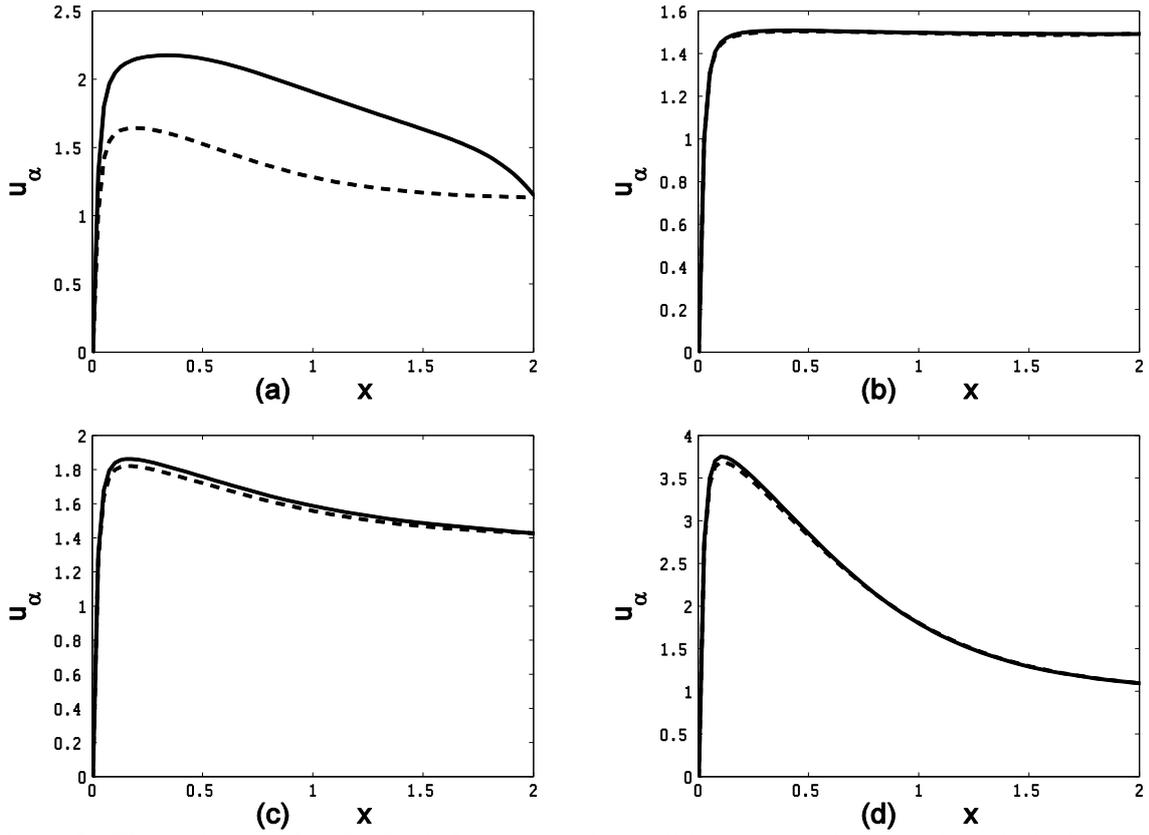

Figure 8. The variation of $u_\alpha$ for R=1, for micropolar (solid curves) and Newtonian (dashed curves), when (a) $\alpha$=-0.5, (b) $\alpha$=0, (c) $\alpha$=0.2, (d) $\alpha$=0.5.

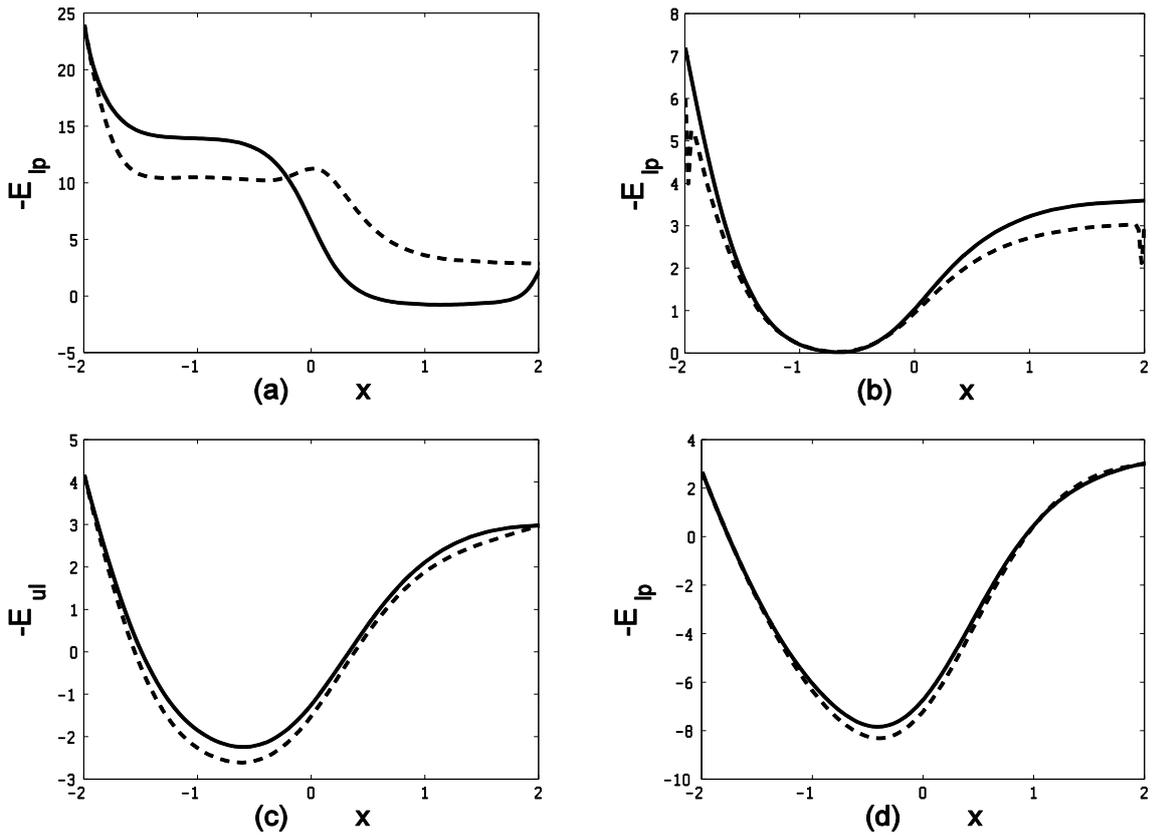

Figure 9. The variation of $-E_\alpha$ for R=1, for a micropolar fluid (solid curves) and a Newtonian fluid (dashed curves), when (a) $\alpha$=-0.5, (b) $\alpha$=0, (c) $\alpha$=0.2, (d) $\alpha$=0.5.

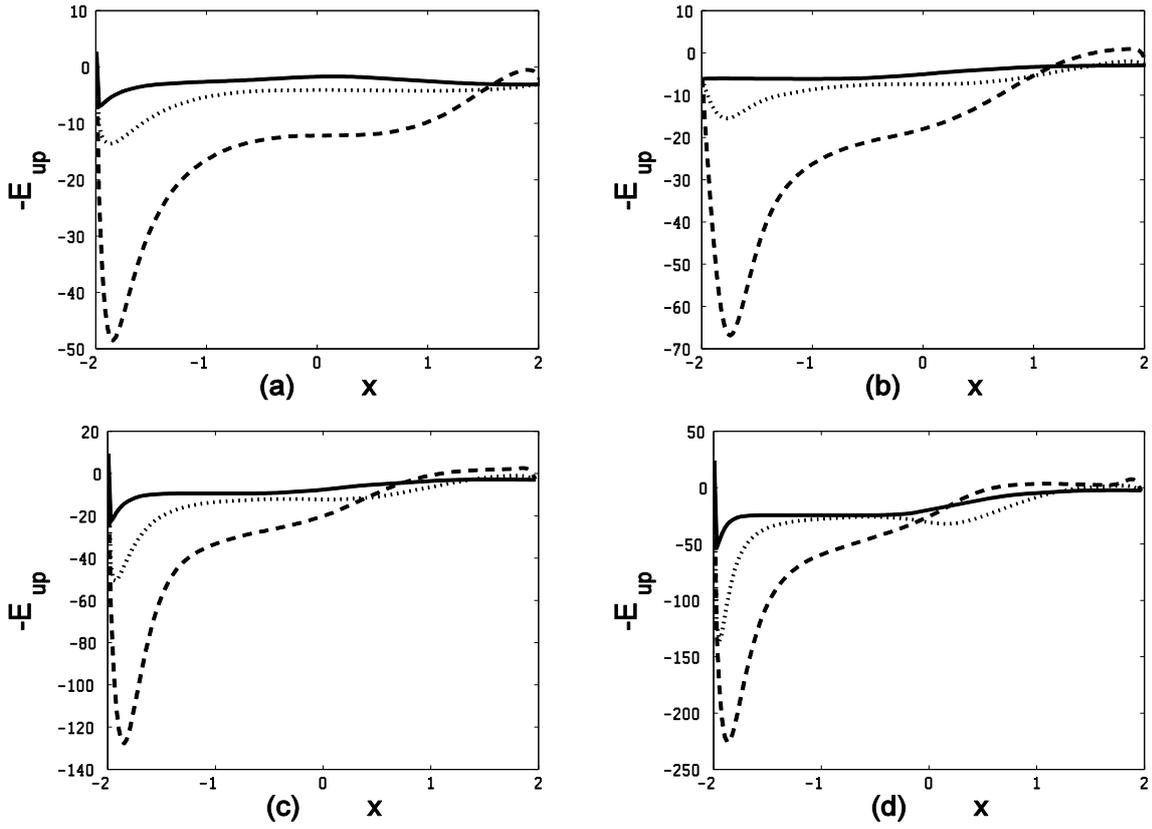

Figure 10. The variation of wall vorticity on the upper plate of channel for R=1 (solid curves), R=100 (dotted curves), and R=1000 (dashed curves), when (a) $\alpha=-0.5$, (b) $\alpha=0$, (c) $\alpha=0.2$, (d) $\alpha=0.5$.

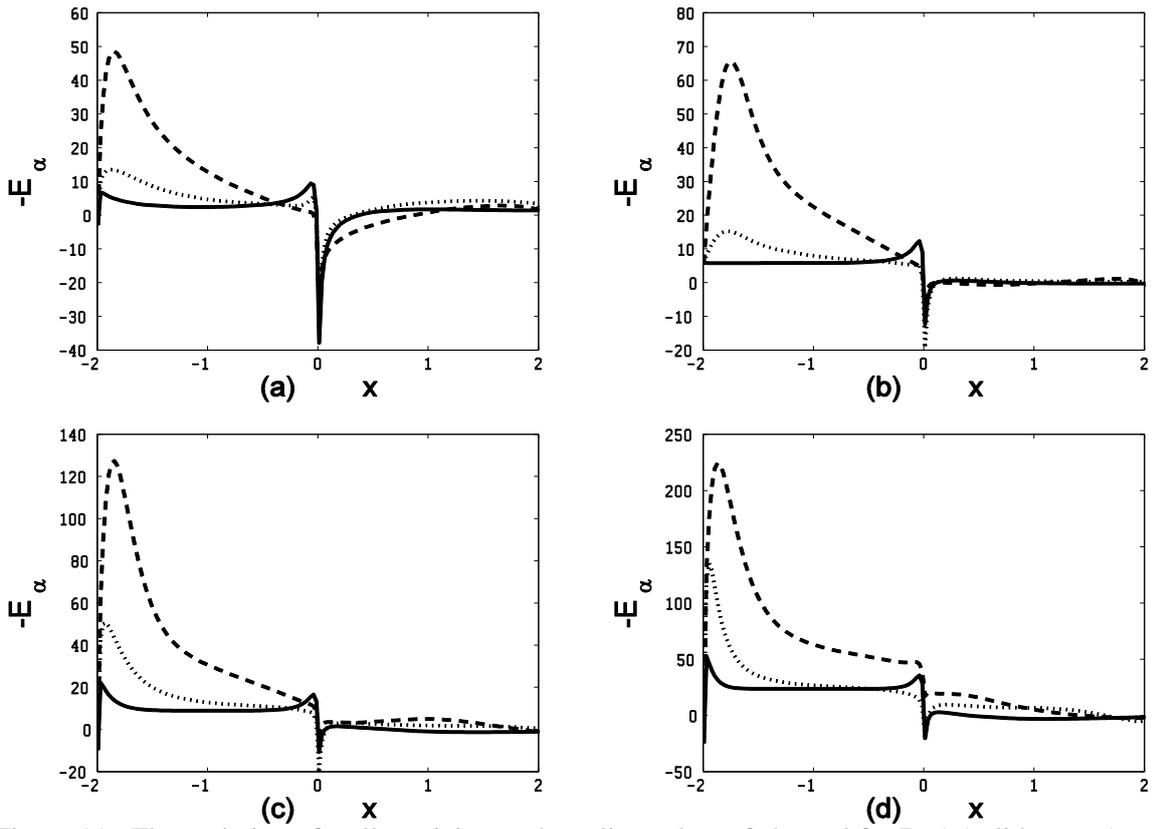

Figure 11. The variation of wall vorticity on the splitter plate of channel for R=1 (solid curves), R=100 (dotted curves), and R=1000 (dashed curves), when (a) $\alpha=-0.5$, (b) $\alpha=0$, (c) $\alpha=0.2$, (d) $\alpha=0.5$.

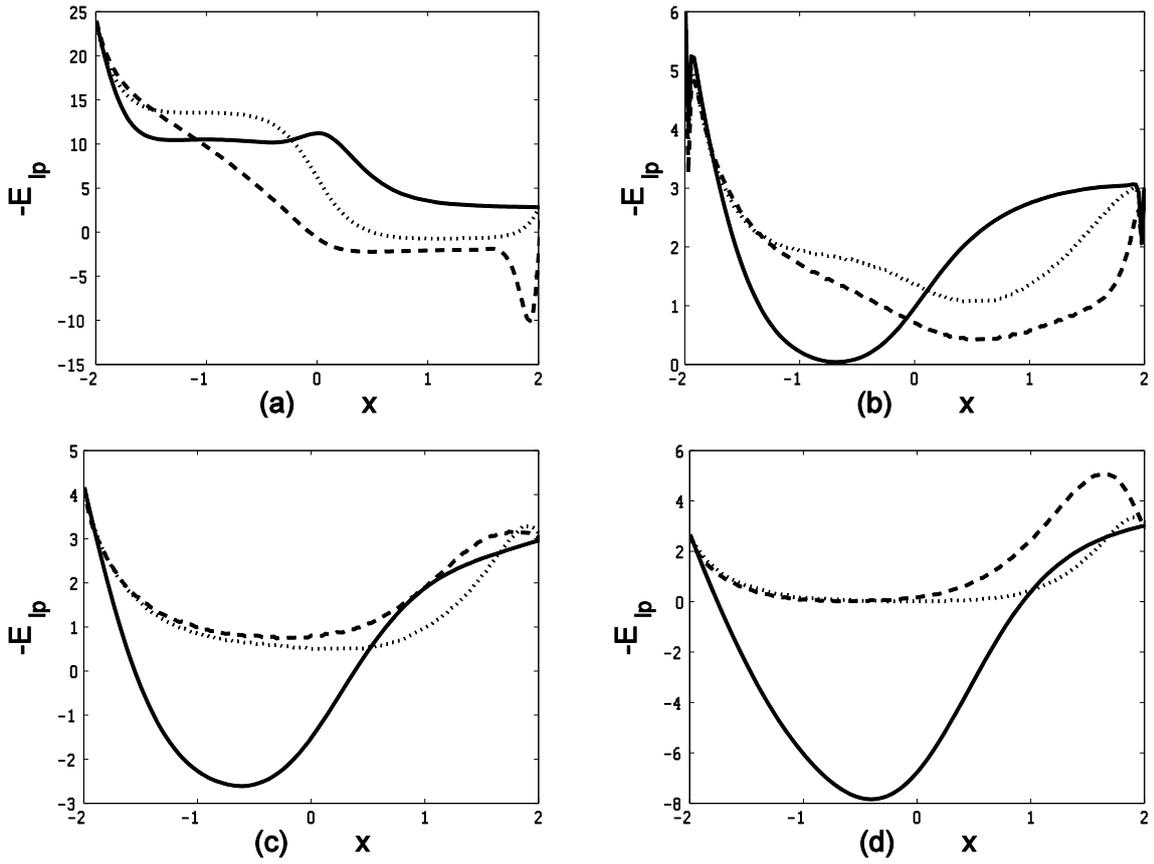

Figure 12. The variation of wall vorticity on the lower plate of channel for R=1 (solid curves), R=100 (dotted curves), and R=1000 (dashed curves), when (a) $\alpha=-0.5$, (b) $\alpha=0$, (c) $\alpha=0.2$, (d) $\alpha=0.5$.